\documentclass[12pt]{article}
\usepackage{amssymb,epsfig}

\renewcommand{\thefootnote}{\fnsymbol{footnote}}

\begin{document}

\title{
\begin{flushright}
\ \\*[-80pt] 
\begin{minipage}{0.2\linewidth}
\normalsize
YITP-06-26 \\
TU-771 \\
KUNS-2028 \\*[50pt]
\end{minipage}
\end{flushright}
{\Large \bf 
Remark on integrating out heavy moduli  
in flux compactification
\\*[20pt]}}

\author{Hiroyuki~Abe$^{1,}$\footnote{
E-mail address: abe@yukawa.kyoto-u.ac.jp}, \ 
Tetsutaro~Higaki$^{2,}$\footnote{
E-mail address: tetsu@tuhep.phys.tohoku.ac.jp} \ and \ 
Tatsuo~Kobayashi$^{3,}$\footnote{
E-mail address: kobayash@gauge.scphys.kyoto-u.ac.jp} \\*[20pt]
$^1${\it \normalsize 
Yukawa Institute for Theoretical Physics, Kyoto University, 
Kyoto 606-8502, Japan} \\
$^2${\it \normalsize 
Department of Physics, Tohoku University, 
Sendai 980-8578, Japan} \\
$^3${\it \normalsize 
Department of Physics, Kyoto University, 
Kyoto 606-8502, Japan} \\*[50pt]}

\date{
\centerline{\small \bf Abstract}
\begin{minipage}{0.9\linewidth}
\medskip 
\medskip 
\small
We study two steps of moduli stabilization in type IIB
flux compactification with gaugino condensations.
We consider the condition that one can integrate out heavy moduli first
 with light moduli remaining.
We give appendix, where detail study is carried out for 
potential minima of the model with a
six dimensional compact space with $h_{1,1}=h_{2,1}=1$, 
including the model, whose respective moduli with 
$h_{1,1} , h_{2,1} \neq 1$ are identified.
\end{minipage}
}

\begin{titlepage}
\maketitle
\thispagestyle{empty}
\end{titlepage}


\renewcommand{\thefootnote}{\arabic{footnote}}
\setcounter{footnote}{0}

\section{Introduction}

Moduli stabilization in superstring theory is 
one of important issues to study.
Indeed, several scenarios have been proposed so far.
Flux compactifications are studied intensively in these years, 
because several moduli can be stabilized through flux 
compactification.
For example, the dilaton $S$ and complex structure moduli $U^\alpha$ 
can be stabilized within the framework of type IIB string 
theory \cite{Giddings:2001yu}, 
while K\"ahler moduli $T$ remain not stabilized.
Recently, in Ref.~\cite{Kachru:2003aw} a new scenario 
was proposed to lead to a de Sitter (or Minkowski) vacuum, 
where all of moduli are stabilized in type IIB string models,
and it is the so-called KKLT scenario.
The KKLT scenario consists of three steps.
At the first step, it is assumed that the dilaton and 
complex structure moduli are stabilized through flux 
compactification.
At the second step, we introduce non-perturbative superpotential 
terms, which depend on the K\"ahler moduli.
That leads to a supersymmetric anti de Sitter (AdS) vacuum.
At the third step, the AdS vacuum is uplifted by 
introducing anti D3 branes, which break supersymmetry (SUSY) 
explicitly.

Phenomenological aspects like soft SUSY breaking terms 
have been studied \cite{Choi:2005ge}.
The KKLT scenario predicts the unique pattern of SUSY breaking 
terms and they have significant phenomenological 
implications \cite{Choi:2005uz}-\cite{Choi:2006bh}.

On the other hand, the flux compactification has been 
studied in explicit models \cite{Kachru:2002he,Lust:2005dy,Curio:2006ea}.
Moreover, the three steps of moduli stabilization has 
been studied, in particular the first two steps.
It has been shown that such two or three steps 
of moduli stabilization may be inconsistent in some models 
showing instability of assumed vacua \cite{Choi:2004sx,deAlwis:2005tf}.

Furthermore, in Ref.~\cite{Abe:2005rx}
models with $S$-$T$ mixing non-perturbative superpotential terms 
have been discussed with the assumption that 
$S$ is already stabilized through flux compactification.
Such models lead to interesting phenomenological and cosmological aspects.
For example, these models have a rich structure of soft SUSY breaking 
terms compared with the original KKLT scenario.
Also, a certain class of these models have moduli potential forms 
different from the original KKLT, and may avoid the 
overshooting problem \cite{Brustein:1992nk} 
and destabilization due to finite temperature 
effects \cite{Buchmuller:2004xr}, from which the original KKLT 
potential suffers.
At any rate, in this new scenario it is the crucial point that 
the one of moduli, say $S$,  
in non-perturbative superpotential is already stabilized through 
the flux compactification.

Thus, it is important to study the validity of 
the two-step moduli stabilization, in particular 
the KKLT type models with moduli-mixing non-perturbative 
superpotential.
That is our purpose of this paper.
Here we concentrate to IIB string models, but 
our discussions  on validity of integrating out heavy 
moduli can be easily extended into generic string theory.

This paper is organized as follows.
In section two, we give a brief review on the KKLT 
scenario and its generalization with moduli-mixing superpotential.
In section three, we study validity of two-steps moduli 
stabilization.
Section four is devoted to conclusion and discussion.
In appendix validity of our procedure is studied 
by examining potential minima explicitly and carefully.

\section{Review on KKLT scenario}

Here we give a brief review on the KKLT scenario 
for moduli stabilization through the flux compactification.
In the KKLT scenario, three types of moduli, the dilaton $S$, 
K\"ahler moduli and complex structure moduli $U^\alpha$ 
are stabilized through two steps.
For simplicity, we consider the string model with a single 
K\"ahler modulus field $T$, although it is straightforward 
to extend our discussions to models with more than one 
K\"ahler moduli.
We use the unit such that $M_{Pl} =1$, where 
$M_{Pl}$ is the 4D reduced Planck mass.

At the first step, we consider a non-trivial background with 
non-vanishing flux, which generates a superpotential 
of $S$ and $U^\alpha$ in type IIB string theory \cite{Gukov:1999ya},
\begin{equation}
W_{flux}(S,U^\alpha) = \int_{M_6} G_3 \, \wedge \, \Omega ,
\end{equation}
where $G_3=F_3^{RR}-2\pi i SH_3^{NS}$ and  $\Omega$ is 
the holomorphic 3-form.
Note that $T$ does not appear in the flux-induced superpotential 
in type IIB string theory.
The K\"ahler potential is written as 
\begin{equation}
K= -\ln(S+\overline{S}) 
-3\ln(T+\overline{T})-\ln(-i\int_{M_6} \Omega \wedge
\overline{\Omega}) .
\end{equation}
The scalar potential in generic supergravity model 
is  written as 
\begin{equation}
V = e^{K}\left(
D_{a}W\overline{D_{b}W}K^{a\overline{b}}-3|W|^2
\right) ,
\end{equation}
where $D_aW=(\partial_a K) W+ \partial_a W$.
Thus, the above superpotential and the  K\"ahler potential 
lead to the following scalar potential,
\begin{eqnarray}
V &=& e^{K}\left(
\sum_{i,j = S, U^\alpha} D_{i}W\overline{D_{j}W}K^{i\overline{j}}
\right),
\label{eq:noscale}
\end{eqnarray}
because of the no-scale form of the K\"ahler potential of $T$.
We obtain the same result, e.g. in models with three moduli 
fields $T^i$.
By this potential, the moduli fields, $S$ and $U^\alpha$, 
except the K\"ahler modulus $T$ can be stabilized at 
the point, $D_SW=D_{U^\alpha} W=0$.

Next, in the second step, the modulus $T$ is stabilized.
That is, in Ref.~\cite{Kachru:2003aw}, a non-perturbative effect 
is assumed to induce the following superpotential,
\begin{equation}
W = w_0 - C e^{-aT},
\end{equation}
where $w_0 =\langle W_{flux}(S,U^\alpha) \rangle_{D_SW=D_{U^\alpha}
  W=0} $.
Such term can be generated by gaugino condensation on 
D7-brane.
Then, the modulus $T$ can be stabilized at 
$D_T W =0$.
It corresponds to 
\begin{equation}
a Re(T) \approx  \ln (C/w_0), 
\end{equation}
when $a Re T \gg 1$.
Its mass is estimated as 
\begin{equation}
m_T \approx a w_0 .
\end{equation}

The above vacuum has the negative energy, i.e., $V= -3 e^K|W|^2 <0$ 
unless $W=0$ at the above point.
To realize a de Sitter (or Minkowski) vacuum, we need 
another step.
To achieve it, 
the uplifting potential,
\begin{equation}
V_{L}= \frac{D}{(T + \bar T)^{n_P}},
\end{equation}
is added in the KKLT scenario.
This uplifting potential slightly shifts the minimum.
This uplifting potential is an explicit SUSY breaking term, and 
the constant $D$ is fine-tuned such as $V +V_L \approx 0$.
That is, the size of $D$ is estimated as $D= {\cal O}(|w_0|^2)$, 
and the SUSY breaking scale is $w_0$.

In Ref.~\cite{Abe:2005rx}, the above scenario has been extended 
into models with non-perturbative superpotential, where $S$ and $T$ 
are mixing,
\begin{equation}
W_{np}=\sum_m C_m e^{-(b_mS+a_mT)}.
\label{W-np}
\end{equation}
These superpotential terms can be generated by e.g. gaugino condensations, 
where corresponding gauge kinetic functions are written as 
linear combinations of $S$ and $T$. 
Such type of moduli-mixing appears in several types of string models, e.g. 
weakly coupled heterotic string models \cite{Choi:1985bz}, 
heterotic M models \cite{Banks:1996ss}, type IIA intersecting D-brane
models and type IIB magnetized D-brane models \cite{Cremades:2002te}.
Here, the exponent constants $a_m$ and $b_m$ can be negative, but 
they must satisfy the condition 
$b_m\langle Re S \rangle +a_m \langle Re T \rangle > 0$.
We assume that $S$ is already stabilized through 
the first step of the flux compactification, 
and that it is frozen in the above superpotential.
That is, the dynamical mode in the above 
superpotential is only $T$.
Its mass is estimated in a way similar to 
the original KKLT scenario.
This type of models lead to interesting aspects from the viewpoint 
of particle phenomenology and cosmology \cite{Abe:2005rx}.

\section{Integrating out heavy moduli}
\label{sec:iohm}

Here, we study mainly on the first two steps of 
moduli stabilization.
As above, in the KKLT scenario, 
stabilization of $T$ and the other moduli is considered separately. 
That is, in the first step $S$  and $U^\alpha$ are stabilized 
(integrated out), 
and in the second step $T$ is stabilized.
Such potential analysis is valid physically 
if the moduli fields $S$ and $U^\alpha$ are much heavier 
than $T$ with the superpotential\footnote{
This point is confirmed in Appendix by examining potential minima 
explicitly and carefully.}, 
\begin{eqnarray}
W &=& W_{flux}+W_{np}. 
\nonumber
\end{eqnarray} 
Hence, let us evaluate masses of moduli fields.
The masses squared of moduli are obtained by the second derivatives 
of the scalar potential, 
\begin{equation}
\left(
\begin{array}{cc}
V_{a \bar b} & V_{ab} \\
V_{\bar a \bar b} & V_{\bar a b}
\end{array}
\right),
\end{equation}
where each entry is obtained at $D_a W =0$ as 
\begin{eqnarray}
V_{a \bar b}\vert_{D_a W =0} & =& (m_0)^2_{a \bar b} + 
(m_1)^2_{a \bar b} + (m_2)^2_{a \bar b} ,\\
V_{ab}\vert_{D_a W =0} & = & (m_1)^2_{a b} + (m_2)^2_{a b},
\end{eqnarray}
with
\begin{eqnarray}
(m_0)^2_{a \bar b} &=& e^K K^{c \bar d} W_{c a} \bar W_{\bar b \bar d} ,\\
(m_1)^2_{a \bar b} &=& e^K \bar W K^{c \bar d} W_{ac}(K_{\bar b \bar d} - 
K_{\bar b} K_{\bar d}) +h.c. ,\\
(m_2)^2_{a \bar b} &=& e^K|W|^2[K^{c \bar d} (K_{c a}-K_aK_c)
(K_{\bar b \bar d} - K_{\bar b}K_{\bar d}) - 3 K_{a \bar b}] ,\\
(m_1)^2_{a b} & = & -e^K\bar W W_{ab} ,\\
(m_2)^2_{a b} & = & -e^K|W|^2 (K_{ab} -K_a K_b).
\end{eqnarray}

We assume that $S, T, U^\alpha = {\cal O}(1)$, and also 
$e^K$ and its derivatives are of ${\cal O}(1)$.
The above second derivatives of scalar potential 
include two types of mass scales.
One is the superpotential mass, $W_{a b}$, 
and the other is supergravity effect, which is represented 
by the gravitino mass $m_{3/2} = e^{K/2} |W|$.
For example, in $V_{a \bar b}\vert_{D_a W =0}$ we 
have 
\begin{equation}
 (m_0)^2_{a \bar b} = {\cal O}(|W_{ab}|^2), \qquad 
(m_1)^2_{a \bar b}  = {\cal O}(|W_{ab}|m_{3/2}), \qquad 
(m_2)^2_{a \bar b} = {\cal O}(m_{3/2}^2) .
\end{equation}
Note that the third term $(m_2)^2_{a \bar b}$ 
appears somehow universally 
for all of moduli fields.
That implies that if 
\begin{equation}
|W_{ab}| \gg |m_{3/2}|,
\end{equation}
the moduli fields corresponding to large superpotential masses 
can be integrated out first.
Furthermore, all of moduli masses satisfy the above condition and 
the determinant of mass matrix is non-vanishing, 
all of masses squared are positive and the SUSY point, 
$D_a W =0$, is stable.\footnote{
See also Ref.~\cite{Blanco-Pillado:2005fn}.}

Now, let us apply the above discussion to the 
flux compactification.
In general, the superpotential $W_{flux}(S,U^\alpha)$ 
induces mass terms of $S$ and $U^\alpha$, and 
those mass scales are naturally of ${\cal O}(M_{Pl})$.
On the other hand, the mass scale of $T$ is 
of ${\cal O}(a m_{3/2})$ in the above model.
Thus, the procedure that first we integrate out  
$S$ and $U^\alpha$  with $T$ remaining, is valid 
when 
\begin{equation}
|W_{ab}| \gg a m_{3/2},
\end{equation}
 for $a,b = S,U^\alpha$.

Here we give two illustrating examples.
The first example is the model without complex structure moduli.
In this model, we obtain 
\begin{equation}
W_{flux} = A + SB,
\end{equation}
where $A$ and $B$ are constants.
This superpotential does not include the mass term, i.e. 
$(W_{flux})_{SS} = 0$.
Thus, the dilaton mass is naturally of the gravitino mass, 
i.e. $m_S={\cal O}(m_{3/2})$.
That is, the dilaton is not heavier than the modulus $T$, 
and it is not valid to integrate out $S$ first by using 
$D_S W_{flux}=0$.
Indeed, it has been shown that it is inconsistent 
to first integrate out $S$ in Ref.~\cite{Choi:2004sx,deAlwis:2005tf}.

The second example is the orientifold model with 
a single complex structure $U$ \cite{Lust:2005dy}.
In this model, we obtain 
\begin{equation}
W_{flux} = A_0 + A_1 U + A_2 S + A_3 SU,
\end{equation}
where $A_i$ ($i=0,1,2,3$ ) are constants.
This superpotential includes a mass term between $S$ and $U$, 
and its natural scale is of ${\cal O}(M_{Pl})$.
Thus, it is valid to integrate out $S$ and $U$ first 
with $T$ remaining if $m_T \approx a m_{3/2} \ll M_{Pl}$.
Note that this mass term has mixing between $S$ and $U$.
That implies that it is not valid to integrate out 
only U by use of $D_U W_{flux} =0$, with $S$ remaining.
Therefore, we have to integrate out $S$  and $U$ at 
the same time.
If the condition $m_T \approx a m_{3/2} \ll M_{Pl}$ is not 
satisfied and $T$ is heavy, we can not integrate out first $S$  and $U$.
Instead of that, we have to study moduli stabilization 
for $S$, $U$ and $T$ at  the same time, and 
the natural order of $m_{3/2}$ is of $M_{Pl}$.\footnote{
In this case, the SUSY breaking scale is $M_{Pl}$, 
even after uplifting to realize de Sitter (or Minkowski) 
vacuum.
That is not good from the phenomenological purpose 
to realize the low-energy SUSY.}

Now, let us consider the condition that we can integrate 
out $S$  and $U$, i.e., $m_T \approx a m_{3/2} \ll M_{Pl}$.
The natural order of 
$w_0 =\langle W_{flux}(S,U^\alpha) \rangle_{D_SW=D_{U^\alpha}
  W=0} \approx m_{3/2} $ is of ${\cal O}(1)$ in the unit 
$M_{pl}=1$.
However, the above condition implies that 
$w_0 =\langle W_{flux}(S,U^\alpha) \rangle_{D_SW=D_{U^\alpha}
  W=0} \approx m_{3/2}\ll {\cal O}(1)$.
One way to realize such condition is to fine-tune 
flux such that $w_0 =\langle W_{flux}(S,U^\alpha) \rangle_{D_SW=D_{U^\alpha}
  W=0} $ is finite, but suppressed compared with $M_{Pl}$.

Another way is to consider the flux compactification 
satisfying 
\begin{eqnarray}
W_{flux} &=& (W_{flux})_a \ = \ 0, 
\label{eq:fnsmin}
\end{eqnarray}
for $a=S,U$.
On top of that, we add non-perturbative term, e.g. 
\begin{eqnarray}
C e^{-bS},
\label{eq:gcond3}
\end{eqnarray}
which can be induced e.g. by gaugino condensation.
The above condition (\ref{eq:fnsmin}) may be rather 
easily realized compared with the condition 
$w_0 =\langle W_{flux}(S,U^\alpha) 
\rangle_{D_SW=D_{U^\alpha} W=0} \approx m_{3/2} \ll {\cal O}(1)$ 
and $w_0 \neq 0$.
We give such an example from Ref.~\cite{Kachru:2002he}, 
which has the following superpotential;
\begin{equation}
W_{flux} = -4(iU^3 +1) +2S(U^3 -3i U^2 - 3 U +2i).
\end{equation}
The SUSY minimum $D_aW_{flux}=0$ corresponds to 
\begin{equation}
U= -i\omega, \qquad S = -i 2 \omega,
\end{equation}
where $\omega = e^{2 \pi i /3}$.
Indeed, this minimum leads to Eq.~(\ref{eq:fnsmin}).

Here we consider the condition leading to 
Eq.~(\ref{eq:fnsmin}) for $S$ and $U$.
We write the flux-induced superpotential 
\begin{equation}
W_{flux} = f^{RR}(U) + S f^{NS}(U),
\end{equation}
where $f^{RR}(U)$ and $f^{NS}(U)$ are polynomial functions of 
$U$.
We write values of $S$ and $U$ at the minimum 
as $S_0$ and $U_0$.
The above conditions requires 
\begin{equation}
f^{RR}(U_0) =  f^{NS}(U_0) =0.
\end{equation}
Thus, we can write 
\begin{equation}
f^{RR}(U) = (U-U_0)^{n_{RR}}\tilde f^{RR}(U), \qquad 
f^{NS}(U) = (U-U_0)^{n_{NS}}\tilde f^{NS}(U),
\end{equation}
with positive integers $n_{RR}$ and $n_{NS}$, where 
$\tilde f^{RR,NS}(U_0) \neq 0$.
Furthermore, the above condition $W_U=0$ at $U_0$ requires 
\begin{equation}
n_{RR}(U-U_0)^{n_{RR}-1}\tilde f^{RR}(U_0) 
+ n_{NR}S_0(U-U_0)^{n_{NS}-1}\tilde f^{NS}(U_0) = 0.
\end{equation}
Obviously, we are interested in the case with $S_0 \neq 0$.
Thus, there are three cases: 1) the case with $n_{RR}= n_{NS} =1$, 
2) the case with $n_{RR}= n_{NS} =2$ and 
3) the case where both $n_{RR}$ and $n_{NS}$ are larger than two, 
i.e., $n_{RR}, n_{NS} \geq 3$. 
In the first case,  the above condition reduces 
\begin{equation}
\tilde f^{RR}(U_0) 
+ S_0\tilde f^{NS}(U_0) = 0,
\end{equation}
that is, $S_0$ is determined as 
\begin{equation}
S_0 = -\frac{\tilde f^{RR}(U_0)}{\tilde f^{NS}(U_0)}.
\end{equation}
Furthermore, since the real part of $S_0$ 
gives the gauge coupling, the obtained value of $S_0$ 
must satisfy $Re(S_0) > 0$.

On the other hand, in the second case with $n_{RR}, n_{NS} = 2$, 
the value $S_0$ is not determined.
Actually, we have 
\begin{equation}
(W_{flux})_{SU} = 0,
\end{equation}
at $U_0$, although we have $(W_{flux})_{UU} \neq 0$ at $U_0$.
That implies that through this type of flux compactification 
only the $U$ moduli is stabilized, but the dilaton $S$ 
is not stabilized. In the third case with 
$n_{RR}, n_{NS} \geq 3$, both the moduli $S$ and $U$ 
are not stabilized by the flux. 

Since in the first case,  $S$ has already a larger mass of ${\cal O}(M_{Pl})$,
the minimum does not shift significantly by adding $C e^{-bS}$ 
as well as terms like Eq.~(\ref{W-np}),
and the added term leads to a small gravitino mass 
$m_{3/2} = e^{K/2}\langle C e^{-bS} \rangle$, which 
is needed to stabilize $T$ at the second stage.
This possibilities has been pointed out in Ref.~\cite{Abe:2005rx}.

Concerned about stabilizing $T$ at the second step,
there is a way not to add the superpotential $C e^{-bS}$ 
to $W_{flux}$, but we change $T$-dependent superpotential 
$W_{np}$. 
We consider not a single term $e^{-aT}$, but more terms like 
\begin{equation}
C_1 e^{-a_1T} - C_2 e^{-a_2T},
\end{equation}
that is, the racetrack model.
In this case, the mass of $T$ is obtained 
\begin{equation}
m_T \approx a_1 a_2 \left( \frac{|C_1|a_1}{|C_2|a_2} 
\right)^{a_1/(a_1  - a_2)}.
\end{equation}
and it can be smaller than $M_{Pl}$.

In the second case with $n_{RR}, n_{NS} = 2$, 
after the first step of the flux compactification, 
two moduli $S$ and $T$ remain light.
Stabilization of such moduli has been discussed 
by non-perturbative superpotential, 
e.g. moduli-mixing racetrack model \cite{Abe:2005pi}, 
which leads to a SUSY breaking vacuum  with negative vacuum 
energy before uplifting.

Here we have studied the model with a single $U$.
The above discussion can be extended to models 
with more than one moduli fields $U^\alpha$.

\section{Conclusion and discussion}

We have studied two steps of moduli stabilization 
through flux compactification and non-perturbative 
superpotential.
We need mass hierarchy between superpotential 
masses $W_{ab}$ and the gravitino mass such that 
the two-step procedure is valid.
Such situation would be realized by fine-tuning 
flux such as $W_{ab} \gg \langle W_{flux} \rangle$, 
although the natural scale of the gravitino mass 
through the flux superpotential would be of 
${\cal O}(M_{Pl})$.
If we do not consider such fine-tuning, 
it would be interesting to use the flux leading to 
$\langle W_{flux} \rangle = 0$.
With this flux, both of $U$ and $S$ are stabilized, 
or only $U$ is stabilized.
Thus, after flux compactification, 
only $T$ modulus remains light, or 
two moduli $T$ and $S$ remain light.
Remaining moduli can be stabilized at the second step.

\subsection*{Acknowledgement}
H.~A.\/,  T.~H.\/ and T.~K.\/ are supported in part by the
Grand-in-Aid for Scientific Research \#182496, \#171643 
and  \#17540251, respectively.
T.~K.\/ is also supported in part by 
the Grant-in-Aid for
the 21st Century COE ``The Center for Diversity and
Universality in Physics'' from the Ministry of Education, Culture,
Sports, Science and Technology of Japan.

\appendix

\section{Perturbation of fluxed no-scale minimum 
by gaugino condensations}
In this appendix, we estimate the shift of the 
potential minimum from the no-scale one 
(\ref{eq:fnsmin}) caused by the gaugino 
condensations, and show that the general 
argument that ``when $S$ and $U$ have heavy masses 
through flux compactification, we can integrate out 
them at the first step with only $T$ remaining,'' 
holds in a concrete and typical situation. 
We assume an effective theory described by 
4D $N=1$ supergravity parameterized by the 
following K\"ahler and superpotential\footnote{
Generalization to the case with more than one 
K\"ahler and complex structure modulus might be 
straightforward.}: 
\begin{eqnarray}
K &=& -n_S \ln (S+\bar{S}) 
-n_T \ln (T+\bar{T}) 
-n_U \ln (U+\bar{U}), 
\label{eq:kahler} \\
W &=& f(S,U)+g(S,T), 
\nonumber
\end{eqnarray}
where the superpotential terms $f(S,U)=W_{flux}$ and 
$g(S,T)=W_{np}$ may originate from the flux and 
gaugino condensations with moduli-mixed gauge couplings, 
respectively, given by 
\begin{eqnarray}
f(S,U) &=& f^{RR}(U)+Sf^{NS}(U), 
\nonumber \\
g(S,T) &=& \sum_m 
C_m e^{-(b_m S+a_m T)}. 
\nonumber
\end{eqnarray}

\subsection{No-scale minimum}
First we analyze a SUSY minimum without 
gaugino condensations, i.e. $W=f(S,U)$. 
In this case, some combinations of flux 
may allow the global SUSY minimum realized by 
conditions $f_S=f_U=f=0$, which result in   
$S=-f^{RR}_U(U)/f^{NS}_U(U)$ and 
$f^{RR}(U)=f^{NS}(U)=0$ 
where $f^{RR,NS}_U=\partial_U f^{RR,NS}$. 
Note that the global SUSY condition $W_a=W=0$ satisfies 
the SUSY stationary condition $D_aW=W_a+K_aW=0$ 
in the supergravity. 
We denote $S$ and $U$ satisfying these conditions 
by $S_0$ and $U_0$, i.e., 
\begin{eqnarray}
\langle U \rangle &=& U_0 
\qquad \textrm{such that} \qquad 
f^{RR}(U_0) \ = \ f^{NS}(U_0) \ = \ 0, 
\nonumber \\
\langle S \rangle &=& S_0 
\ = \ -f^{RR}_U(U_0)/f^{NS}_U(U_0).
\label{eq:nsv}
\end{eqnarray}
Note that K\"ahler modulus $T$ remains as 
a flat direction in the case with $n_T=3$ 
for which the scalar potential is in the 
no-scale form (\ref{eq:noscale}), and we 
assume $n_T=3$ in the following arguments. 

The moduli masses $m^2_{a\bar{b}}$, 
$m^2_{ab}=\overline{m^2_{\bar{a}\bar{b}}}$ are evaluated by 
computing the second derivatives of the scalar potential, 
\begin{eqnarray}
m^2_{a\bar{b}} &=& 
K_{a\bar{a}}^{-1/2}K_{b\bar{b}}^{-1/2} V_{a\bar{b}}, 
\nonumber \\
m^2_{ab} &=& 
K_{a\bar{a}}^{-1/2}K_{b\bar{b}}^{-1/2} V_{ab} .
\nonumber
\end{eqnarray}
In the case with $W=f(S,U)$, we find 
\begin{eqnarray}
V_{a\bar{b}} \Big|_{S=S_0,U=U_0} &=& 
e^K K^{c\bar{d}} f_{ca} \bar{f}_{\bar{d}\bar{b}}, 
\nonumber \\
V_{ab} \Big|_{S=S_0,U=U_0} &=& 0. 
\label{eq:susymass}
\end{eqnarray}
The moduli fields $S$ and $U$ typically receive 
heavy masses if $f_{ab} \ne 0 $ ($a=S,U$), 
because the parameters in the flux superpotential 
$f(S,U)$ are expected to be naturally of ${\cal O}(1)$
in the unit $M_{Pl}=1$. 

\subsection{Including gaugino condensations}
Next we consider the case with $W=f(S,U)+g(S,T)$ and 
study the perturbation of the previous SUSY vacuum 
caused by gaugino condensations described 
by additional superpotential terms in $g(S,T)$. 
For such purpose, we analyze the shift of the vacuum 
\begin{eqnarray}
T &=& T_0+\delta T, 
\nonumber \\
S &=& S_0+\delta S, 
\nonumber \\
U &=& U_0+\delta U, 
\label{eq:fullvac}
\end{eqnarray}
around the vacuum satisfying 
$D_S f=D_U f=0$ and $D_T g=0$ defined by 
\begin{eqnarray}
f \big|_0 &=& f_S \big|_0 \ = \ f_U \big|_0 \ = \ 0, 
\label{eq:noscalegc} \\
g \big|_0 &=& -g_T/K_T \big|_0 \ \ne \ 0,
\nonumber
\end{eqnarray}
where $\big|_0$ stands for $\big|_{T=T_0,S=S_0,U=U_0}$. 
The first derivatives of $G=K+\ln|W|^2$ can be expanded as 
\begin{eqnarray}
G_A &=& G_A \big|_0 
+\delta \phi^B G_{AB} \big|_0 
+{\cal O}(\delta^2), 
\nonumber
\end{eqnarray}
where the indices $A,B,C,\ldots$ run all the holomorphic and 
anti-holomorphic fields as 
$A,B,C,\ldots=(S$, $T$, $U$, $\bar{S}$, $\bar{T}$, $\bar{U})$ and 
$\delta \phi^A$ denotes the deviation of the vacuum value 
$\delta \phi^A=\delta S$, $\delta T$, $\delta U$, 
$\delta \bar{S}$, $\delta \bar{T}$, $\delta \bar{U}$ 
with the corresponding index. 

At the linear order of $\delta \phi^A$, the solution 
of the SUSY stationary condition $G_A=0$ is given by 
\begin{eqnarray}
\delta \phi^A &=& -G^{AB} G_B \big|_0 
+{\cal O}(\delta^2), 
\label{eq:dphi}
\end{eqnarray}
where $G_{AB}G^{BC}=\delta_A^{\ C}$. 
If we assume that all the parameters in $f(S,U)$ and 
$g(S,T)$ are of order one quantities except for 
$a_m \sim b_m \gg 1$, we may naturally obtain 
\begin{eqnarray}
K \big|_0,\ K_A \big|_0,\ K_{AB} \big|_0,\ 
\ldots &\sim& {\cal O}(1), 
\nonumber \\
f_{ab} \big|_0,\ f_{abc} \big|_0,\ f_{abcd} \big|_0,\ 
\ldots &\sim& {\cal O}(1), 
\nonumber \\
g \big|_0 \ \ll \ g_{ab} \big|_0 \ \ll \ g_{abc} \big|_0,\ 
\ldots &\ll& 1, 
\nonumber
\end{eqnarray}
if nonvanishing, by which we can expect the hierarchical structure 
\begin{eqnarray}
G_{UU} \big|_0,\ G_{SU} \big|_0,\ \gg \ 
G_{SS} \big|_0,\ G_{ST} \big|_0,\ G_{TT} \big|_0,\ \gg \ 
G_{a\bar{b}} \big|_0=K_{a\bar{b}} \big|_0,\ \gg \ 
G_{TU} \big|_0=0, 
\label{eq:hiergij}
\end{eqnarray}
where $G_{ab}=K_{ab}+W_{ab}/W-(W_a/W)(W_b/W)$. 
From this, we can approximate $G_{AB}$ by 
the block-diagonal form, 
\begin{eqnarray}
G_{AB} \big|_0 &\sim& 
\left( 
\begin{array}{cc}
G_{ab} \big|_0 & 0 \\ 
0 & G_{\bar{a}\bar{b}} \big|_0
\end{array}
\right), 
\nonumber
\end{eqnarray}
and the same for its inverse $G^{AB}$. 
This means that Eq.~(\ref{eq:dphi}) becomes 
a holomorphic equation, 
$\delta \phi^a=-G^{ab}G_b \big|_0 +{\cal O}(\delta^2)$, 
and with the explicit form of $G^{ab}$ we find 
\begin{eqnarray}
\left( \begin{array}{c} 
\delta T \\ 
\delta S \\ 
\delta U 
\end{array} \right) &=& 
g \times \left. \left( \begin{array}{c} 
-\frac{G_{ST}}{G_{TT}} \frac{1}{f_{SU}}
\left( \frac{f_{UU}}{f_{SU}}G_S-G_U \right) \\ 
\frac{1}{f_{SU}}
\left( \frac{f_{UU}}{f_{SU}}G_S-G_U \right) \\ 
-\frac{1}{f_{SU}} G_S 
\end{array} \right) \right|_0 
+{\cal O}(g^2), 
\label{eq:dtdsdu}
\end{eqnarray}
where 
\begin{eqnarray}
G_T \big|_0 &=& 
{\cal G}_T \big|_0 \ = \ 0, 
\nonumber \\
G_S \big|_0 &=& 
{\cal G}_S \big|_0 \ = \ (g_S/g+K_S) \big|_0 
\ \sim \ {\cal O}(b_m) 
\ \sim \ {\cal O}(a_m), 
\nonumber \\
G_U \big|_0 &=& 
{\cal G}_U \big|_0 \ = \ K_U \big|_0 
\ \sim \ {\cal O}(1), 
\label{eq:gi0}
\end{eqnarray}
and ${\cal G} \equiv K+\ln |g|^2$.
Therefore we find 
$\delta \Phi^a/\Phi^a_0 \sim {\cal O}(g) \ll 1$ 
with $\Phi^a=(T,S,U)$ for 
$\Phi^a_0 = \Phi^a \big|_0 \sim {\cal O}(1)$, 
and this linearized analysis is enough to study 
the combined effect of the fluxes and the gaugino 
condensations.

\subsection{Moduli mass}
Here we estimate the moduli masses 
and mixing at the vacuum (\ref{eq:fullvac}) 
determined by Eq.~(\ref{eq:dtdsdu}). 
Up to the second order of $\delta \phi^i$ 
we can expand the moduli masses as\footnote{
The coefficient $K_{AA}^{-1/2} K_{BB}^{-1/2}$ 
in $m^2_{AB}$ originates from the normalization 
of kinetic terms. Note that here we are assuming 
the K\"ahler potential without moduli mixing 
(\ref{eq:kahler}).} 
\begin{eqnarray}
m^2_{AB} &=& 
K_{AA}^{-1/2} K_{BB}^{-1/2} V_{AB} 
\nonumber \\ &=& 
m^2_{AB} \big|_0 
+\delta \phi^C\, \partial_C m^2_{AB} \big|_0 
+\frac{1}{2} \delta \phi^C \delta \phi^D\, 
\partial_C \partial_D m^2_{AB} \big|_0 
+{\cal O}(\delta^3) 
\nonumber \\ &=& 
Z_{AB}^{(2)} V_{AB} \big|_0 
+\delta \phi^C\, Z_{AB}^{(1)} V_{ABC} \big|_0 
+\frac{1}{2} \delta \phi^C \delta \phi^D\, 
Z_{AB}^{(0)} V_{ABCD} \big|_0 
+{\cal O}(\delta^3), 
\nonumber
\end{eqnarray}
where 
\begin{eqnarray}
Z_{AB}^{(2)} &=& Z_{AB}^{(1)}
-\frac{1}{2} Z_{AB}^{(0)} \Big\{ 
K_{AA}^{-1}K_{AA,CD}+K_{BB}^{-1}K_{BB,CD}
\nonumber \\ && \qquad 
-\frac{3}{2} \big( K_{AA}^{-2}K_{AA,C}K_{AA,D}
+K_{BB}^{-2}K_{BB,C}K_{BB,D} \big) 
\nonumber \\ && \qquad 
-\frac{1}{2} \big( K_{AA}^{-1}K_{BB}^{-1}K_{AA,C}K_{BB,D}
+K_{AA}^{-1}K_{BB}^{-1}K_{AA,D}K_{BB,C} \big) 
\Big\} \delta \phi^C \delta \phi^D, 
\label{eq:z2} \\
Z_{AB}^{(1)} &=& 
Z_{AB}^{(0)} \Big\{ 1
-\frac{1}{2}\big( K_{AA}^{-1}K_{AA,C}
+K_{BB}^{-1}K_{BB,C} \big) \delta \phi^C \Big\}, 
\label{eq:z1} \\
Z_{AB}^{(0)} &=& K_{AA}^{-1/2} K_{BB}^{-1/2}. 
\nonumber
\end{eqnarray}
All the derivatives of scalar potential 
$V_A$, $V_{AB}$, $\ldots$ can be 
written in terms of $G=K+\ln|W|^2$ and 
its derivatives, $G_A$, $G_{AB}$, $\ldots$. 
At the point (\ref{eq:noscalegc}), these can be written as 
\begin{eqnarray}
G \big|_0 &=& {\cal G} \big|_0, \quad 
G_A \big|_0 \ = \ {\cal G}_A \big|_0, \quad 
G_{AB} \ = \ {\cal G}_{AB} \big|_0 + g^{-1} f_{AB} \big|_0, \quad 
\ldots, 
\nonumber
\end{eqnarray}
where ${\cal G}=K+\ln |g|^2$. 
By calculating $V_{AB}$, $V_{ABC}$, $V_{ABCD}$ 
and taking Eq.~(\ref{eq:hiergij}) into account, 
we find the leading contributions to 
each component in moduli mass matrices squared as 
\begin{eqnarray}
e^{-K}m^2_{a\bar{b}} &\sim& 
\left. \left( \begin{array}{c|cc}
(\mu_{T\bar{T}}+\delta \mu_{T\bar{T}})\,|g|^2 & 
(\mu_{T\bar{S}}+\delta \mu_{T\bar{S}})\,|g|^2 & 
\delta \mu_{T\bar{U}} \,g  \\*[5pt] \hline 
\Big( \overline{\mu_{T\bar{S}}}
+\overline{\delta \mu_{T\bar{S}}} \Big)\,|g|^2 & 
K_{S\bar{S}}^{-1}K_{U\bar{U}}^{-1}|f_{SU}|^2 & 
K_{S\bar{S}}^{-1/2}K_{U\bar{U}}^{-3/2} 
f_{SU} \bar{f}_{\bar{U}\bar{U}} \\*[5pt]
\overline{\delta \mu_{T\bar{U}}}\,\bar{g} & 
K_{S\bar{S}}^{-1/2}K_{U\bar{U}}^{-3/2} 
\bar{f}_{\bar{S}\bar{U}} f_{UU} & 
K_{U\bar{U}}^{-1}(K_{S\bar{S}}^{-1}|f_{SU}|^2 
+K_{U\bar{U}}^{-1}|f_{UU}|^2) 
\end{array} \right) \right|_0, \qquad 
\label{eq:mijb} \\*[10pt]
e^{-K}m^2_{ab} &\sim& 
\left. \left( \begin{array}{c|cc}
(\mu_{TT}+\delta \mu_{TT})\,|g|^2 & 
(\mu_{TS}+\delta \mu_{TS})\,|g|^2 & 
\delta \mu_{TU}\,|g|^2 \\*[5pt] \hline 
(\mu_{TS}+\delta \mu_{TS})\,|g|^2 & 
(\mu_{SS}+\delta \mu_{SS})\,|g|^2 & 
\delta \mu_{SU}\,\bar{g} \\*[5pt]
\delta \mu_{TU}\,|g|^2 & 
\delta \mu_{SU}\,\bar{g} & 
\delta \mu_{UU}\,\bar{g} 
\end{array} \right) \right|_0, 
\label{eq:mij}
\end{eqnarray}
where the row and column correspond to $a,b=(T,S,U)$. 

The coefficients $\mu_{AB}$ represent the contributions 
from ${\cal G}$, 
\begin{eqnarray}
\mu_{AB} &=& 
K_{AA}^{-1/2} K_{BB}^{-1/2} e^{-{\cal G}} {\cal V}_{AB}, 
\qquad 
{\cal V} \ = \ e^{\cal G} \big( 
{\cal G}^{a\bar{b}}{\cal G}_a{\cal G}_{\bar b} -3\big), 
\nonumber
\end{eqnarray}
and given by, e.g., 
\begin{eqnarray}
\begin{array}{rclcl}
\mu_{T\bar{T}} 
&=& K_{T\bar{T}}^{-1}
(K_{T\bar{T}}^{-1}|{\cal G}_{TT}|^2 
+K_{S\bar{S}}^{-1}|{\cal G}_{ST}|^2) 
-2+K_{S\bar{S}}^{-1}|{\cal G}_S|^2 
&\sim& {\cal O}(a_m^4), \\*[5pt]
\mu_{TT} 
&=& (K_{T\bar{T}}K_{S\bar{S}})^{-1}( 
|{\cal G}_S|^2 {\cal G}_{TT} 
+\bar{\cal G}_{\bar{S}}{\cal G}_{TTS}) 
-{\cal G}_{TT} 
&\sim& {\cal O}(a_m^4), 
\end{array}
\nonumber
\end{eqnarray}
and similarly 
$\mu_{T\bar{S}}$, $\mu_{SS}$, $\mu_{TS} \sim {\cal O}(a_m^4)$. 
On the other hand, the terms with coefficients 
$\delta \mu_{AB}$ come from the remaining contributions 
($g$-$f$ mixed terms in $V_{AB}$, $V_{ABC}$, $V_{ABCD}$), 
and are given by, e.g., 
\begin{eqnarray}
\begin{array}{rclcl}
\delta \mu_{T\bar{T}} &=& 
-K_{S\bar{S}}^{-1}|{\cal G}_S|^2 
&\sim& {\cal O}(a_m^2), \\*[5pt] 
\delta \mu_{T\bar{U}} 
&=& (K_{T\bar{T}}K_{U\bar{U}})^{-1/2}K_{S\bar{S}}^{-1} 
{\cal G}_{ST} \bar{f}_{\bar{S}\bar{U}} 
&\sim& {\cal O}(a_m^2), \\*[5pt]
\delta \mu_{TT} &=& 
-(K_{T\bar{T}}K_{S\bar{S}})^{-1}( 
|{\cal G}_S|^2 {\cal G}_{TT} 
+\bar{\cal G}_{\bar{S}}{\cal G}_{TTS}) 
&\sim& {\cal O}(a_m^4), 
\end{array}
\nonumber
\end{eqnarray}
and similarly 
$\delta \mu_{TS}$, 
$\delta \mu_{TU}$, 
$\delta \mu_{SS}$, 
$\delta \mu_{SU}$, 
$\delta \mu_{UU} \le {\cal O}(a_m^4)$. 

First, from the $2 \times 2$ sub-matrix in Eq.~(\ref{eq:mijb}), 
$m^2_{a\bar{b}}$ with $a,b=(S,U)$, we find that 
the moduli $S$ and $U$ generically receive ${\cal O}(M_{Pl})$ of heavy 
masses. 
Second, since
${{\cal O}(a_m^4)}$ of contributions 
in $\mu_{TT}$ and $\delta \mu_{TT}$ cancel each other, 
$\mu_{TT}+\delta \mu_{TT}=-{\cal G}_{TT} \sim {\cal O}(a_m^2)$ 
while 
$\mu_{T\bar{T}}+\delta \mu_{T\bar{T}} \sim {\cal O}(a_m^4)$, 
we find $m^2_{T\bar{T}} \gg m^2_{TT}$ and 
the imaginary direction, ${\rm Im}\,T$ is not destabilized.

So far, to estimate $\mu_{TT},\mu_{T\bar{T}}$ and their 
deviations,  
we have considered the case with $g(S,T)$, where 
two or more $a_m$ are non-vanishing.
When only a single $a_m$ is non-vanishing and $g(S,T)$ includes 
other $S$-dependent terms as well as constant term, 
the above estimation would change because of e.g. 
${\cal G}_{TT} = {\cal O}(a_m)$.
However,  the imaginary direction, ${\rm Im}\,T$ is stable still.

The third point is that, 
the $T$-$U$ mixing $m^2_{T\bar{U}} \sim {\cal O}(a_m^2 |g|)$ can 
affect the lightest eigenvalue of $m^2_{a\bar{b}}$ 
that is ${\cal O}(a_m^4 |g|^2)$. 
If we normalize $m^2_{a \bar{b}}$ and define 
the $3 \times 3$ matrix\footnote{
Here we assume $f_{SU} \ne 0$, and omit 
$m^2_{T\bar{S}}$, $m^2_{S\bar{T}} \sim |g|^2$ 
which does not affect the following order estimation 
when $f_{SU} \ne 0$.} 
\begin{eqnarray}
{\cal M} &\equiv& 
e^{-K}K_{S \bar{S}}K_{U \bar{U}} 
|f_{SU}|^{-2} m^2_{a \bar{b}} 
\ = \ \left( 
\begin{array}{ccc}
Z|g|^2 & 0 & Y\,g \\*[5pt]
0 & 1 & X \\*[5pt]
\bar{Y}\,\bar{g} & \bar{X} & 1+|X|^2 
\end{array}
\right), 
\nonumber
\end{eqnarray}
where 
\begin{eqnarray}
X &=& (K_{S \bar{S}}/K_{U \bar{U}})^{1/2} 
(f_{\bar{U} \bar{U}}/f_{\bar{S} \bar{U}}), 
\nonumber \\
Y &=& (K_{U \bar{U}}/K_{T \bar{T}})^{1/2} 
({\cal G}_{ST}/f_{SU}), 
\nonumber \\
Z &=& 
K_{S \bar{S}}K_{U \bar{U}} |f_{SU}|^{-2} 
\{ K_{T\bar{T}}^{-1}
(K_{T\bar{T}}^{-1}|{\cal G}_{TT}|^2 
+K_{S\bar{S}}^{-1}|{\cal G}_{ST}|^2) -2 \}, 
\nonumber
\end{eqnarray}
the eigenvalue equation for ${\cal M}$ is given by 
\begin{eqnarray}
-\lambda^3+(2+|X|^2+Z|g|^2)\lambda^2
-\big( 1+\{ Z(2+|X|^2)-|Y|^2 \}|g|^2 \big) \lambda 
+(Z-|Y|^2)|g|^2 &=& 0, 
\nonumber
\end{eqnarray}
where $\lambda$ is the eigenvalue of ${\cal M}$. 
For the lightest eigenvalue $\lambda \sim {\cal O}(|g|^2)$, 
this equation is approximated as 
\begin{eqnarray}
-\lambda+(Z-|Y|^2)|g|^2 +{\cal O}(|g|^4) &=& 0, 
\nonumber
\end{eqnarray}
and we find the mass squared of the lightest mode ${\cal T}$ as 
\begin{eqnarray}
m^2_{{\cal T}\bar{\cal T}} &\equiv& 
e^K (K_{S \bar{S}}K_{U \bar{U}})^{-1} |f_{SU}|^2 \lambda 
\ = \ e^K (K_{T\bar{T}}^{-2}|{\cal G}_{TT}|^2 -2)|g|^2 \Big|_0. 
\nonumber
\end{eqnarray}  
This is actually the same as the mass squared 
$m^2_{{\cal T}\bar{\cal T}}$ calculated from 
the generalized K\"ahler potential 
\begin{eqnarray}
G({\cal T}) &=& 
K(S_0,{\cal T},U_0) + \ln \big| g(S_0,{\cal T}) \big|^2, 
\label{eq:effgk}
\end{eqnarray}
at the SUSY point $G_{\cal T}({\cal T})=0$, 
which supports the fact that the low energy effective theory 
of the light mode ${\cal T}$ is described by $G({\cal T})$. 
In addition, large eigenvalues of ${\cal M}$ are obtained as 
\begin{equation}
\lambda = \frac{1}{2}\left( 
2+|X|^2 \pm |X| \sqrt{4+|X|^2}\right) + O(|g|^2).
\end{equation}
These coincide with masses squared, which are obtained from 
(\ref{eq:susymass}) and are positive.

We finally comment that all the analyses and results 
in this appendix would be applied to the perturbation 
of (fine-tuned) AdS minimum 
\begin{eqnarray}
f \big|_0 &=& 
-f_S/K_S \big|_0 \ = \ -f_U/K_U \big|_0 
\ \equiv \ w_0 \ \sim \ {\cal O}(g|_0), 
\nonumber \\
W \big|_0 &=& w_0+g \big|_0 \ = \ -g_T/K_T \big|_0, 
\nonumber
\end{eqnarray}
instead of the no-scale minimum (\ref{eq:noscalegc}). 
This will be done by replacing 
$g|_0 \to w_0+g|_0$ everywhere,  
$g(S_0,{\cal T}) \to w_0+g(S_0,{\cal T})$ in Eq.~(\ref{eq:effgk}) 
and forgetting Eq.~(\ref{eq:gi0}), 
at least as long as the following condition holds 
\begin{eqnarray}
f_a \big|_0 &=& -K_a \big|_0 w_0 
\ \sim \ {\cal O}(g|_0) 
\ \ll \ g_b \big|_0 \ \sim {\cal O}(a_m g_0), 
\nonumber
\end{eqnarray}
for $a=(S,U)$ and $b=(S,T)$. 
However, as shown in Sec.~\ref{sec:iohm}, a (fine-tuned) 
nonvanishing value of $w_0 \sim {\cal O}(g|_0)$ is not 
necessary in order to stabilize ${\cal T}$ through $G({\cal T})$. 
To do that, we can assume, e.g., the existence of (\ref{eq:gcond3}) 
in $g(S,T)$ which generates a constant superpotential term 
$C e^{-bS_0} \sim {\cal O}(g|_0)$ naturally and effectively 
in $G({\cal T})$.

\end{document}